\documentclass[12pt,british]{article}
\pdfoutput=1
\usepackage[T1]{fontenc}
\usepackage[latin9]{inputenc}
\usepackage{geometry}
\usepackage{color}
\usepackage{amsmath}
\usepackage{amssymb}
\usepackage{stackrel}
\usepackage{graphicx}
\usepackage{esint}
\usepackage[numbers]{natbib}

\makeatletter

\providecommand{\tabularnewline}{\\}

\numberwithin{equation}{section}
\numberwithin{figure}{section}
\numberwithin{table}{section}

\usepackage{babel}
\pdfoutput=1

\@ifundefined{showcaptionsetup}{}{%
 \PassOptionsToPackage{caption=false}{subfig}}
\usepackage{subfig}
\makeatother

\usepackage{babel}
\begin{document}

\title{Overlaps after quantum quenches in the sine-Gordon model}

\author{D. X. Horváth$^{1,2}$ and G. Takács$^{1,2}$~\\
 $^{1}${\small{}MTA-BME \textquotedbl{}Momentum\textquotedbl{} Statistical
Field Theory Research Group}\\
 {\small{}1111 Budapest, Budafoki út 8, Hungary}\\
 $^{2}${\small{}Department of Theoretical Physics, }\\
{\small{} Budapest University of Technology and Economics}\\
{\small{}1111 Budapest, Budafoki út 8, Hungary} }

\date{3rd April 2017}
\maketitle
\begin{abstract}
We present a numerical computation of overlaps in mass quenches in
sine-Gordon quantum field theory using truncated conformal space approach
(TCSA). To improve the cut-off dependence of the method, we use a
novel running coupling definition which has a general applicability
in free boson TCSA. The numerical results are used to confirm the
validity of a previously proposed analytical Ansatz for the initial
state in the sinh-Gordon quench. 
\end{abstract}

\section{Introduction}

One of the most challenging problems in contemporary physics is the
understanding of dynamical and relaxation phenomena in closed quantum
systems out of equilibrium. Motivated by both theoretical interest
and experimental relevance, recent studies led to a series of interesting
discoveries such as the experimental observation of the lack of thermalization
in integrable systems \citep{NewtonCradle,ExperimentalNoThermalization1,ExperimentalNoThermalization2,ExperimentalNoThermalization3}.
To explain the stationary state of integrable quantum systems, the
concept of the generalized Gibbs ensemble (GGE) was proposed \citep{GGEProposal},
and recently experimentally confirmed \citep{GGEExperimental}. It
also turned out that the GGE was generally incomplete when only including
the well-known local conserved charges \citep{CauxNoGGE,PozsgayNoGGE},
and its completion made necessary the inclusion of novel quasi-local
charges \citep{ProsenCGGE,IlievskiQuasiLocal}. Adding to this the
unconventional, often ballistic nature of quantum transport \citep{BallisticTransport,ProsenTransport}
or the confinement effects in the spread of correlations in non-integrable
systems \citep{Confinement} indeed, a remarkable range of exotic
behaviour has emerged in recent years. 

A paradigmatic framework for non-equilibrium dynamics is provided
by quantum quenches \citep{CardyCalabrese}, in which the initial
state (which is typically the ground state of some pre-quench Hamiltonian
$H_{0}$) is subject to evolution driven by a post-quench Hamiltonian
$H$, which is obtained from $H_{0}$ by instantaneously changing
some parameters of the system. For the purpose of computing the time
evolution it is useful to know the overlaps, i.e. the amplitudes of
the post-quench excitations in the initial state. Indeed, in the case
of integrable post-quench dynamics, knowledge of these overlaps often
enables the determination of steady state properties, and even the
time evolution \citep{ExcitedBetheStates,FiorettoMussardo,QuenchAction,SineGSemiClassical,DeLuca}.
However, the determination of the overlaps is generally a very difficult
task. When both the pre-quench and post-quench theories are non-interacting,
the overlaps can be determined using the Bogoliubov transformation
linking the pre- and post-quench excitation modes, but in genuinely
interacting integrable models there are only few cases in which the
overlaps are explicitly known. These cases mostly include spin chains
and the Lieb-Liniger model \citep{PozsiKozlowski,PozsiXXZOverlaps,XXZGaudinDet,PiroliXXZ,BrockmannOverlapNeel,NardisLL}. 

Quantum field theories are known to provide universal descriptions
of statistical models and many-body systems, valid at long distances,
and therefore quantum quenches in field theories are interesting,
especially in the quest for universal characteristics and behaviour
under quantum quenches. In massive relativistic integrable quantum
field theories there exists a number of efficient approaches to the
quench dynamics, which depend on the assumption that the initial state
$|\Psi(0)\rangle$ can be written in a squeezed vacuum form in terms
of post-quench Zamolodchikov-Faddeev creation operators $Z_{a}^{\dagger}(\vartheta)$
for asymptotic particle states and the post-quench vacuum $|0\rangle$

\begin{equation}
|\Psi(0)\rangle=\mathcal{N}\exp\int\frac{d\vartheta}{2\pi}K_{a,b}(\vartheta)Z_{a}^{\dagger}(-\vartheta)Z_{b}^{\dagger}(\vartheta)|0\rangle\ ,\label{eq:IntegrableQuench}
\end{equation}
which is just the analogue of the Bogoliubov solution for free theories.
The above form of the initial state is equivalent to the statement
that the multi-particle creation amplitudes factorize into products
of independent single pair creation amplitudes. This is obviously
reminiscent of the factorisation property of scattering in integrable
quantum field theories \citep{ZamZam}, which justifies calling this
class of quenches ``integrable''. Such a form of the initial state
enables the application of methods based on thermodynamic Bethe Ansatz
(TBA) \citep{ExcitedBetheStates,FiorettoMussardo,QuenchAction}, form
factor based spectral expansions \citep{SchurichtEssler,BertiniSineG}
or semi-classical approach \citep{SineGSemiClassical}. However, even
within the class of integrable quenches no exact solutions are known
for the overlap functions $K_{a,b}$ apart from non-interacting quantum
field theory models. 

Recently an Ansatz for the overlaps was proposed for the quench from
a massive free boson to an interacting sinh-Gordon model \citep{SotiriadisTakacsMussardo,InitalStateIntEqHierarchcy},
which has already been used to obtain predictions for steady state
expectation values \citep{BertiniPiroliCalabrese}. The aim of the
present work is to provide a test of this solution from first principles,
by comparing their analytical continuation to sine-Gordon theory to
a direct evaluation of the overlaps in the framework of the Truncated
Conformal Space Approach (TCSA), originally introduced in \citep{YurZam}
and extended to the sine-Gordon model in \citep{c1TCSA}.

\section{Overlaps in quantum field theory quenches\label{sec:TheoryOfOverlaps}}

\subsection{The sinh-Gordon quench overlaps and their continuation to sine-Gordon }

The work \citep{InitalStateIntEqHierarchcy} considered a quench from
a massive free bosonic theory with particle mass $m_{0}$ to the massive
sinh-Gordon theory 
\begin{equation}
\mathcal{A}=\int d^{2}x\left(\frac{1}{2}\partial_{\mu}\Phi\partial^{\mu}\Phi-\frac{\mu^{2}}{g^{2}}\cosh g\Phi\right)\:,\label{eq:SinhGAction}
\end{equation}
with coupling $g$ and physical particle mass $m$ (which is equal
to $\mu$ in the classical limit). For these quenches it was argued
that the initial state can be cast into the exponential form 

\begin{equation}
|\Psi(0)\rangle=\mathcal{N}\exp\left\{ \int\frac{d\vartheta}{2\pi}K(\vartheta)Z^{\dagger}(-\vartheta)Z^{\dagger}(\vartheta)\right\} |0\rangle\ .\label{eq:ExpInitStateShG}
\end{equation}
However, properly demonstrating that the initial state has the form
\eqref{eq:IntegrableQuench} in terms of post-quench asymptotic particle
states is far from straightforward, and has only been possible in
the non-interacting case (also in some interacting quenches in spin
chains and Bose gases, where the exact overlaps are known and factorize
in the thermodynamic limit \citep{PozsiKozlowski,PozsiXXZOverlaps,XXZGaudinDet,PiroliXXZ,BrockmannOverlapNeel,NardisLL}). 

A class of states which has the exponential form is given by so-called
integrable boundary states introduced in \citep{GoshalZamo}; however,
they cannot be considered physical initial states since they are not
normalizable. As shown in \citep{CardyCalabrese,Cardy_further} it
is possible to construct proper initial states in terms of a boundary
state $|B\rangle$ using the form
\[
|\Psi(0)\rangle=e^{-\sum\tau_{i}Q_{i}}|B\rangle
\]
where the $Q_{i}$ are local conserved charges. Assuming that (as
usual) the one-particle states are eigenstates of the $Q_{i}$, this
obviously preserves the exponential form of the state, but it is hard
to identify the physical quench (i.e. the pre-quench Hamiltonian $H_{0}$)
that results in this state for a particular choice of the real parameters
$\tau_{i}$.

For quenches starting from a large initial mass in the sinh-Gordon
field theory arguments in favour of the exponential form were advanced
in \citep{SotiriadisTakacsMussardo}, and even the following Ansatz
was proposed for the function $K$:
\begin{eqnarray}
K(\vartheta) & = & K_{free}(\vartheta)K_{D}(\vartheta)\:,\label{eq:Ansatz}
\end{eqnarray}
where $K_{free}(\vartheta)$ is given by 
\[
K_{free}(\vartheta)=\frac{E_{0}(\vartheta)-E(\vartheta)}{E_{0}(\vartheta)+E(\vartheta)},\quad E(\vartheta)=m\cosh\vartheta\quad,\quad E_{0}(\vartheta)=\sqrt{m_{\text{0}}^{2}+m^{2}\sinh^{2}\vartheta}\:,
\]
and is identical (up to a sign) with the Bogoliubov amplitude in a
mass quench $m_{0}\longrightarrow m$ within a free bosonic model,
while $K_{D}(\vartheta)$ is the amplitude of the Dirichlet boundary
($\Phi=\text{0}$) state in sinh-Gordon theory:\footnote{$K_{D}$ can be obtained by analytic continuation from the first breather
boundary amplitude in sine-Gordon theory which was obtained in \citep{Ghoshal}.} 
\begin{equation}
K_{D}(\vartheta)=i\tanh(\vartheta/2)\frac{\cosh\left(\vartheta/2-i\pi B/8\right)\sinh(\vartheta/2+i\pi(B+2)/8)}{\sinh\left(\vartheta/2+i\pi B/8\right)\cosh(\vartheta/2-i\pi(B+2)/8)}\quad,\quad B(g)\,=\,\frac{2g^{2}}{8\pi+g^{2}}\:.\label{eq:KD}
\end{equation}
In the follow-up work \citep{InitalStateIntEqHierarchcy} it was shown
that provided the initial state contains only multiple particle states
composed of pairs with opposite momenta, extensivity of the charges
guaranteeing integrability leads to factorisation of multi-pair amplitudes
and therefore results in an exponential form of the state, the only
undetermined parameter being the pair creation amplitudes $K_{a,b}(\vartheta)$.
However, the pair structure itself remains mainly an assumption supported
only by some heuristic arguments \citep{InitalStateIntEqHierarchcy}. 

Furthermore an infinite integral equation hierarchy was derived that
determines (at least in principle) the full form of the initial state
in terms of the post-quench multi-particle states for the quenches
from a free massive boson to the the sinh-Gordon model, and it was
further shown that the simple Ansatz \eqref{eq:Ansatz} was a very
good numerical solution of the lowest member of the hierarchy provided
the exponential form of the initial state was assumed. In addition,
the next member of the hierarchy was used for a numerical test of
the factorisation assumption itself, which worked well within the
limitations of the numerics. 

Continuing to imaginary couplings $g=i\beta$ results in sine-Gordon
theory 
\begin{equation}
\mathcal{A}=\int d^{2}x\left(\frac{1}{2}\partial_{\mu}\Phi\partial^{\mu}\Phi+\frac{\mu^{2}}{\beta^{2}}\cos\beta\Phi\right)\:,\label{eq:SineGAction-1}
\end{equation}
and it is useful to introduce $\xi=\beta^{2}/(8\pi-\beta^{2})=-B/2$.
The fundamental excitations are a doublet of soliton/antisoliton of
mass $M$. In the attractive regime ($\xi<1$) the spectrum also contains
breathers $B_{r}$ (soliton-antisoliton bound states) with masses
$m_{r}=2M\sin r\pi\xi/2$ with $r$ a positive integer less than $\xi^{-1}$.
Due to integrability, the exact factorized $S$ matrix is also known
\citep{ZamZam}. Under the analytic continuation to imaginary couplings
the sinh-Gordon particle corresponds to the first breather $B_{1}$,
which can be supported both by perturbation theory and the correspondence
between the respective $S$ matrix amplitudes. As a result, form factors
of local operators and reflection factors containing only the first
breather $B_{1}$ are also identical to the corresponding sinh-Gordon
quantities under the same analytic continuation, which are all known
in the model.

Here we consider sine-Gordon quenches which correspond to abruptly
changing the soliton mass $M_{0}\rightarrow M$ while leaving the
interaction strength $\xi$ unaltered in the Hamiltonian $H$ associated
with \eqref{eq:SineGAction-1}. Note that under the analytic continuation
this is related to a mass quench within sinh-Gordon theory with a
fixed coupling $g$, while the Ansatz \eqref{eq:Ansatz} was obtained
for a quench from a free boson to sinh-Gordon theory. However, provided
the interaction in the initial Hamiltonian does not play a significant
role, we can expect that an analytic continuation 
\begin{align}
K_{B_{1}B_{1}}(\vartheta) & =\frac{E_{0}(\vartheta)-E(\vartheta)}{E_{0}(\vartheta)+E(\vartheta)}K_{D}(\vartheta)\label{eq:sGAnsatz}\\
K_{D}(\vartheta) & =i\tanh\left(\frac{\vartheta}{2}\right)\frac{\cosh\left(\frac{\vartheta}{2}+\frac{i\pi\xi}{4}\right)}{\sinh\left(\frac{\vartheta}{2}-\frac{i\pi\xi}{4}\right)}\frac{\sinh\left(\frac{\vartheta}{2}+\frac{i\pi(1-\xi)}{4}\right)}{\cosh\left(\frac{\vartheta}{2}-\frac{i\pi(1-\xi)}{4}\right)}\nonumber 
\end{align}
gives a good approximation to the first breather pair creation amplitude
in the sine-Gordon mass quench. We shall return to the issue of the
initial interaction later when discussing the numerical results. Note
that the amplitude depends only on the quench mass ratio, which is
the same for each particle species since $\xi$ is fixed, so we substituted
the first breather mass by the soliton mass.

\subsection{\label{sub:Overlaps-in-finite} Overlaps in finite volume}

In our numerical calculation we consider the system in a finite volume
$L$ with periodic boundary conditions, therefore we briefly recall
the theory of finite size dependence of boundary state amplitudes,
worked out in \citep{OnePointFunctions}. To keep the formulas short,
we consider only one species of particles as the generalization to
more than one species is rather obvious. Denoting the pre-quench ground
state by $|B\rangle$, the most general expansion in terms of the
post-quench eigenstates is

\begin{equation}
|B\rangle=|0\mathcal{\rangle}+\sum_{n=1}^{\infty}\int\prod_{i=1}^{n}\frac{d\ \vartheta_{i}}{2\pi}K_{n}(\vartheta_{1},...\vartheta_{n})\delta\left(\sum_{i=1}^{n}m\sinh\vartheta_{i}\right)|\vartheta_{1},...,\vartheta_{n}\rangle\ ,\label{eq:BExpansionInfVol}
\end{equation}
while in finite volume one obtains

\begin{equation}
|B\rangle_{L}=|0\mathcal{\rangle}_{L}+\sum_{n=1}^{\infty}\sum_{I_{1},...I_{n}}'N_{n}K_{n}(\vartheta_{1}^{*},...\vartheta_{n}^{*})|I_{1},...,I_{n}\rangle_{L}\ ,\label{eq:BExpansionFinVol}
\end{equation}
where $\{\vartheta_{i}^{*}\}$ are the solutions of the Bethe-Yang
equations, i.e. the system

\begin{equation}
Q_{i}=mL\sinh\vartheta_{i}+\sum_{j=1,\neq i}^{n}\delta(\vartheta_{i}-\vartheta_{j})=2\pi I_{i}\ ,i=1,...,n,\label{eq:BYQ}
\end{equation}
where $\delta(\vartheta)=-i\ln S(\vartheta)$ is the phase-shift corresponding
to the two particle $S$-matrix $S(\vartheta)$, $m$ is the physical
mass of the particle and $I_{i}$ are the quantum numbers that characterize
the finite volume states, with the prime meaning that only zero momentum
states are included. In \citep{OnePointFunctions} the $N_{n}$ functions
were explicitly determined 

\begin{equation}
N_{n}(\vartheta_{1}^{*},...\vartheta_{n}^{*})=\frac{\sqrt{\rho_{n}(\vartheta_{1}^{*},...\vartheta_{n}^{*})}}{\bar{\rho}_{n-1}(\vartheta_{1}^{*},...\vartheta_{n-1}^{*})}+O(e^{-\mu'L})\ ,\label{eq:NFunctions}
\end{equation}
where $\rho_{n}$ is the density of states given by the Bethe-Yang
Jacobi determinant \citep{FinVolFFs1}
\[
\rho_{n}=\det\left\{ \frac{\partial Q_{k}}{\partial\vartheta_{j}}\right\} _{j,k=1,\dots,n}
\]
whereas $\bar{\rho}_{n-1}$ is the so-called reduced density of states
which takes into account momentum conservation and is computed as
the Jacobian 
\[
\bar{\rho}_{n-1}=\det\left\{ \frac{\partial\bar{Q}_{k}}{\partial\vartheta_{j}}\right\} _{j,k=1,\dots,n-1}
\]
of the constrained Bethe-Yang equations 
\begin{align}
\bar{Q}_{i} & =mL\sinh\vartheta_{i}+\sum_{j=1,\neq i}^{n-1}\delta(\vartheta_{i}-\vartheta_{j})+\delta(\vartheta_{i}-\tilde{\vartheta})=2\pi I_{i}\ ,i=1,...,n-1\ ,\label{eq:BYQConstZeroMom}\\
 & \tilde{\vartheta}=-\sinh^{-1}(\sum_{i=1}^{n-1}\sinh\vartheta_{i})\nonumber 
\end{align}
Formula \eqref{eq:NFunctions} is exact to all orders in the inverse
volume $L^{-1}$ as indicated by correction terms that decay exponentially
with the volume with some characteristic scale $\mu'$ (cf. \citep{FinVolFFs1}).

For the case when the expansion of the initial state in terms of the
post-quench eigenstates only contains paired states
\begin{equation}
|B\rangle=|0\mathcal{\rangle}+\sum_{n=1}^{\infty}\int\prod_{i=1}^{n}\frac{d\ \vartheta_{i}}{2\pi}K_{n}(\vartheta_{1},...\vartheta_{n})|-\vartheta_{1},\vartheta_{1}...,-\vartheta_{n},\vartheta_{n}\rangle\ ,\label{eq:BExpansionInfVolPaired}
\end{equation}
the appropriate constrained Bethe-Yang equations are 

\begin{equation}
\bar{Q}_{i}^{p}=mL\sinh\vartheta_{i}+\delta(2\vartheta_{i})+\sum_{j\neq i}\delta(\vartheta_{i}-\vartheta_{j})+\delta(\vartheta_{i}+\vartheta_{j})=2\pi I_{i}\ ,i=1,...,n\ ,\label{eq:BYQConstPairs}
\end{equation}
with solution $\{\vartheta_{i}^{*}\}$, and the finite volume expansion
is
\begin{equation}
|B\rangle_{L}=|0\mathcal{\rangle}_{L}+\sum_{n=1}^{\infty}\sum_{I_{1},...I_{n}}N_{n}(\vartheta_{1}^{*},...\vartheta_{n}^{*})K_{n}(\vartheta_{1}^{*},...\vartheta_{n}^{*})|-I_{1},I_{1}...,-I_{n},I_{n}\rangle_{L}\ ,\label{eq:BExpansionFinVolPaired}
\end{equation}
with the $N_{n}$ functions \citep{OnePointFunctions}
\begin{equation}
N_{n}(\vartheta_{1}^{*},...\vartheta_{n}^{*})=\frac{\sqrt{\rho_{2n}(-\vartheta_{1}^{*},\vartheta_{1}^{*}...-\vartheta_{n}^{*},\vartheta_{n}^{*})}}{\bar{\rho}_{n}^{p}(\vartheta_{1}^{*},...\vartheta_{n}^{*})}+O(e^{-\mu'L})\quad,\quad\bar{\rho}_{n}^{p}=\det\left\{ \frac{\partial\bar{Q}_{k}^{p}}{\partial\vartheta_{j}}\right\} _{j,k=1,\dots,n}\:.\label{eq:NFunctionsPaired}
\end{equation}

\section{Overlaps from TCSA}

\subsection{TCSA for the sine-Gordon mass quench}

We now turn to studying sine-Gordon mass quenches in truncated conformal
space approach (TCSA), following the ideas in \citep{Tibi} which
applied a similar approach to Ising field theory. For sine-Gordon,
TCSA consists of representing the model as a compactified free massless
boson conformal field theory (CFT) perturbed by a relevant operator,
with the Hamiltonian

\begin{align}
H & =\int dx\frac{1}{2}:\left(\partial_{t}\Phi\right)^{2}+\left(\partial_{x}\Phi\right)^{2}:-\frac{\lambda}{2}\int dx\left(V_{1}+V_{-1}\right)\label{eq:pcft_Hamiltonian}\\
 & V_{a}=:e^{ia\beta\Phi}:\nonumber 
\end{align}
where the semicolon denotes normal ordering in terms of the massless
scalar field modes. In a finite volume $L$, the spectrum of the free
boson CFT is discrete and can be truncated to a finite subspace by
introducing an upper cut-off $e_{cut}$ in terms of the eigenvalue
of the dilatation operator (which gives the energy in conformal units).
Physical energies and volumes can be expressed in units of the soliton
mass using the relation
\begin{equation}
\lambda=\frac{2\Gamma(\Delta)}{\pi\Gamma(1-\Delta)}\left(\frac{\sqrt{\pi}\Gamma\left(\frac{1}{2-2\Delta}\right)M}{2\Gamma\left(\frac{\Delta}{2-2\Delta}\right)}\right)^{2-2\Delta}\qquad,\qquad\Delta=\frac{\beta^{2}}{8\pi}\label{eq:mass_scale}
\end{equation}
so that the dimensionless volume variable and Hamiltonian can be defined
as $l=ML$ and $h=H/M$, respectively. For more details on the sine-Gordon
TCSA the interested reader is referred to \citep{c1TCSA}.

The initial state corresponds to the ground state of the same Hamiltonian
\eqref{eq:pcft_Hamiltonian} with $\lambda$ replaced by $\lambda_{\text{0}}$
corresponding to $M_{0}$. When considering the post-quench evolution
in dimensionless volume $l=ML$, implementing the quench means using
the ground state computed in the rescaled volume $l_{0}=M_{0}l/M$
\citep{Tibi}.

The cut-off dependence of TCSA can be (partially) eliminated using
renormalisation group methods \citep{TCSARG1,TCSARG2,RGTCSAWatts}.
Here we used a modified version of the running coupling prescription
in \citep{LencsesTakacs}. We can write an effective Hamiltonian in
the form
\[
H_{eff}=\int dx\frac{1}{2}:\left(\partial_{t}\Phi\right)^{2}+\left(\partial_{x}\Phi\right)^{2}:+\lambda_{0}\mathbb{I}+\frac{\lambda_{1}}{2}\int dx\left(V_{1}+V_{-1}\right)+\frac{\lambda_{2}}{2}\int dx\left(V_{2}+V_{-2}\right)
\]
where we included counter terms generated at leading order according
to the fusion rules $V_{a}V_{b}\sim V_{a+b}$. Introducing the dimensionless
couplings
\[
\tilde{\lambda}_{a}=\frac{\lambda_{a}L^{2-2h_{a}}}{(2\pi)^{1-2h_{a}}}\quad h_{a}=\frac{a^{2}\beta^{2}}{8\pi}
\]
the running couplings $\tilde{\lambda}_{i}$ are determined by the
RG equations
\begin{equation}
\tilde{\lambda}_{c}(n)-\tilde{\lambda}_{c}(n-1)=\frac{1}{2n-d_{0}(l)}\sum_{a,b}\tilde{\lambda}_{a}(n)\tilde{\lambda_{b}}(n)C_{ab}^{c}\frac{n^{2h_{abc}-2}}{\Gamma(h_{abc})^{2}}\left(1+O(1/n)\right)\label{eq:RG_eqs}
\end{equation}
where $n$ is the cut-off expressed in conformal levels, $C_{ab}^{c}$
is the operator product coefficient, $h_{abc}=h_{a}+h_{b}-h_{c}$
and $d_{0}(l)$ is the vacuum scaling function (cf. \citep{LencsesTakacs}).
At the lowest order it is only necessary to run the couplings $\lambda_{0}$
and $\lambda_{2}$, from their starting values $\lambda_{0}=0$ and
$\lambda_{2}=0$ at $n=\infty$. 

The couplings must be run following \eqref{eq:RG_eqs} down from $n=\infty$
to the appropriate value of $n_{cut}$ corresponding to the given
cut-off $e_{cut}$. It must be taken into account that the $c=1$
Hilbert space is spanned by Fock modules $\mathcal{F}_{a}$ created
from the vacuum by $V_{a}$ and the Hamiltonian is block-diagonal
in terms of the Fock modules, symbolically:

\[
\left(\begin{array}{ccccccc}
H_{0}+\mathbb{I} & V_{1} & V_{2}\\
V_{-1} & H_{0}+\mathbb{I} & V_{1} & V_{2}\\
V_{-2} & V_{-1} & H_{0}+\mathbb{I} & V_{1} & V_{2}\\
 & \ddots & \ddots & \ddots & \ddots & \ddots\\
 &  & V_{-2} & V_{-1} & H_{0}+\mathbb{I} & V_{1} & V_{2}\\
 &  &  & V_{-2} & V_{-1} & H_{0}+\mathbb{I} & V_{1}\\
 &  &  &  & V_{-2} & V_{-1} & H_{0}+\mathbb{I}
\end{array}\right)
\]
and the eventual value of $n_{cut}$ depends on the block one considers.
Namely, when computing the coefficient of the block $V_{2}$ between
\emph{$\mathcal{F}_{a}$ }and $\mathcal{F}_{a+2}$, the intermediate
states in the OPE $V_{1}V_{1}\sim V_{2}$ are from $\mathcal{F}_{a+1}$
which determines the level $n_{cut}$ appropriate for the given block,
and similarly for $V_{-2}$ between \emph{$\mathcal{F}_{a}$ }and
$\mathcal{F}_{a-2}$ $n_{cut}$ is fixed from $\mathcal{F}_{a-1}$.
For the identity term between \emph{$\mathcal{F}_{a}$ }and \emph{$\mathcal{F}_{a}$}
there are two possible intermediate modules $\mathcal{F}_{a\pm1}$,
so the identity coupling must be split into two pieces $\lambda_{0\pm}$,
each of them running down to the appropriate $n_{cut}$ determined
by the highest level in $\mathcal{F}_{a\pm1}$. 

The block-dependent running coupling corresponds to including a non-local
counter term. The fact that such counter terms are necessary was noted
in \citep{TCSARychkov}; they account for $1/n$ corrections in the
running coupling. In the sine-Gordon there is a large $1/n$ effect
resulting from the fact that the cut-off level is heavily module dependent,
ranging from $e_{cut}$ in Fock module $\mathcal{F}_{0}$ to $0$
for the Fock modules with the largest indices $\mathcal{F}_{\pm a_{max}}$.
The consistency of this scheme was verified by numerically checking
the cut-off dependence of the $15$ lowest-lying levels in the TCSA
spectrum, which proved to be negligible with this method.

\subsection{The $B_{1}-B_{1}$ pair amplitude}

Now we turn to numerical results for the $B_{1}-B_{1}$ pair amplitudes
and compare them with the infinite volume prediction \eqref{eq:sGAnsatz}.
The first task is to identify states corresponding to $B_{1}-B_{1}$
pairs in the numerical spectrum of the post-quench Hamiltonian. Solving
the constrained Bethe-Yang system \eqref{eq:BYQConstPairs} one obtains
the possible rapidities from which the energy levels can be computed.
However, apart from the lowest lying levels, the identification is
not feasible by merely comparing the TCSA and the Bethe-Yang energies
due the density of the TCSA spectrum. This issue can be overcome by
supplementing the energy selection procedure with a comparison of
the finite volume form factors of the fields $V_{1}$ and $V_{2}$
obtained using the formalism developed in \citep{FinVolFFs1}, to
the TCSA matrix elements (for an exposition of how this works in sine-Gordon
theory cf. \citep{FeherTakacs}). As the form factors depend sensitively
on the particle content of the state, the identification can unambiguously
be performed.

\begin{figure}[th]
\subfloat[$R=2.3$, $M/M_{0}=0.5$, $M_{0}L=55$, $e_{cut}=24$]{\begin{centering}
\includegraphics[width=0.45\columnwidth]{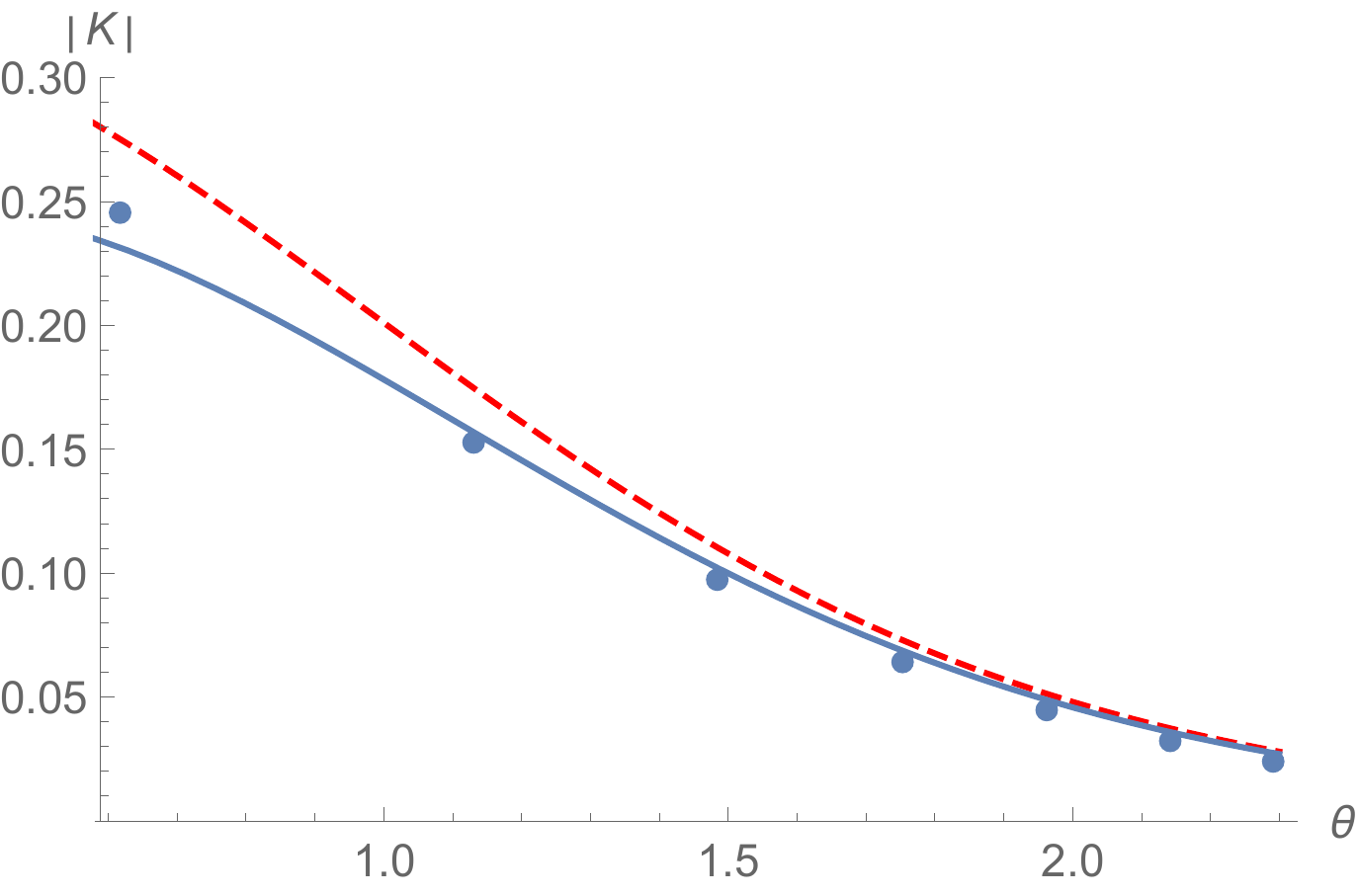}
\par\end{centering}

}\subfloat[$R=2.3$, $M/M_{0}=0.75$, $M_{0}L=40$, $e_{cut}=22$]{\begin{centering}
\includegraphics[width=0.45\textwidth]{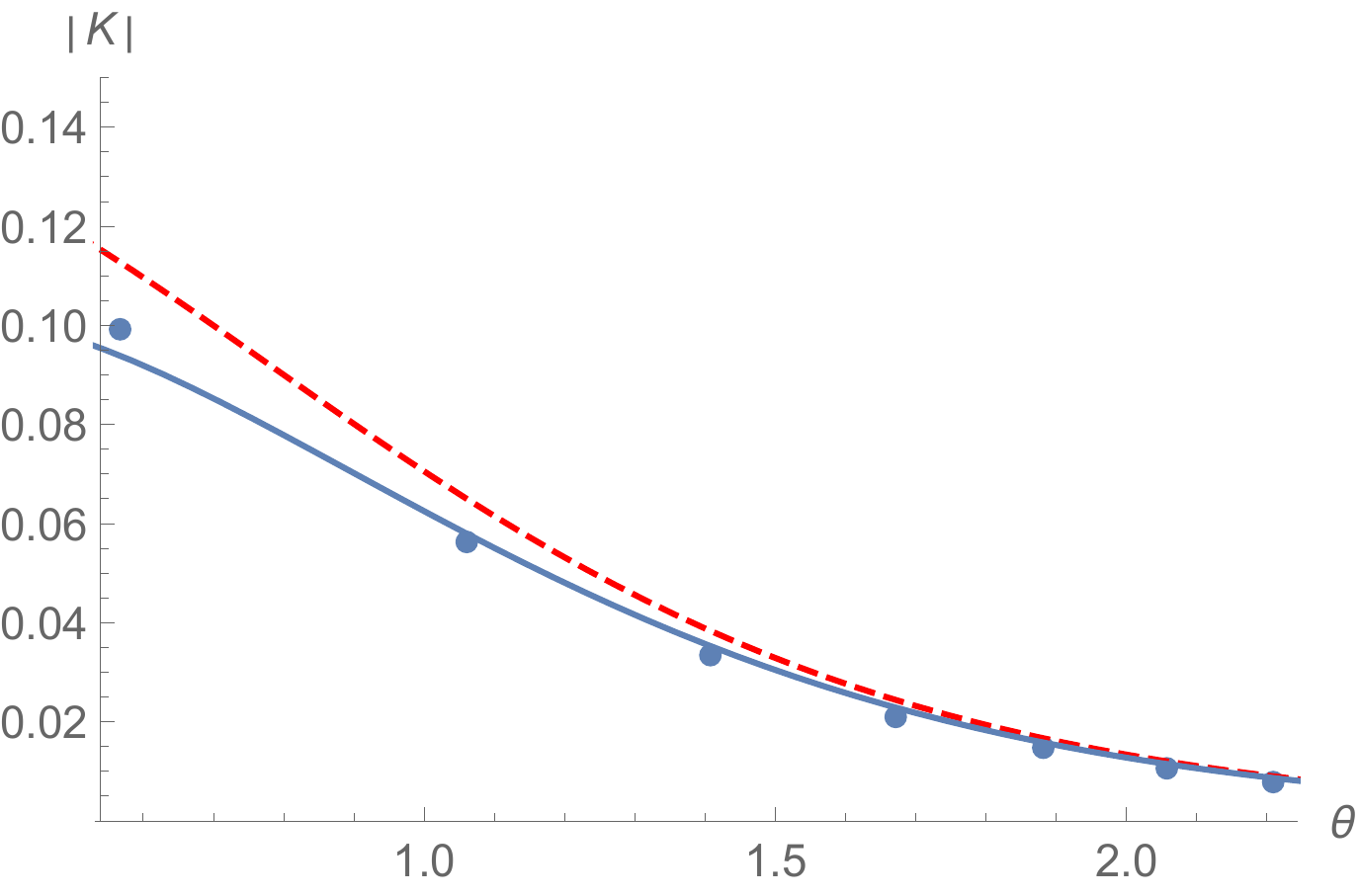}
\par\end{centering}

}\\

\subfloat[$R=2.3$, $M/M_{0}=1.5$, $M_{0}L=22$, $e_{cut}=24$]{\begin{centering}
\includegraphics[width=0.45\columnwidth]{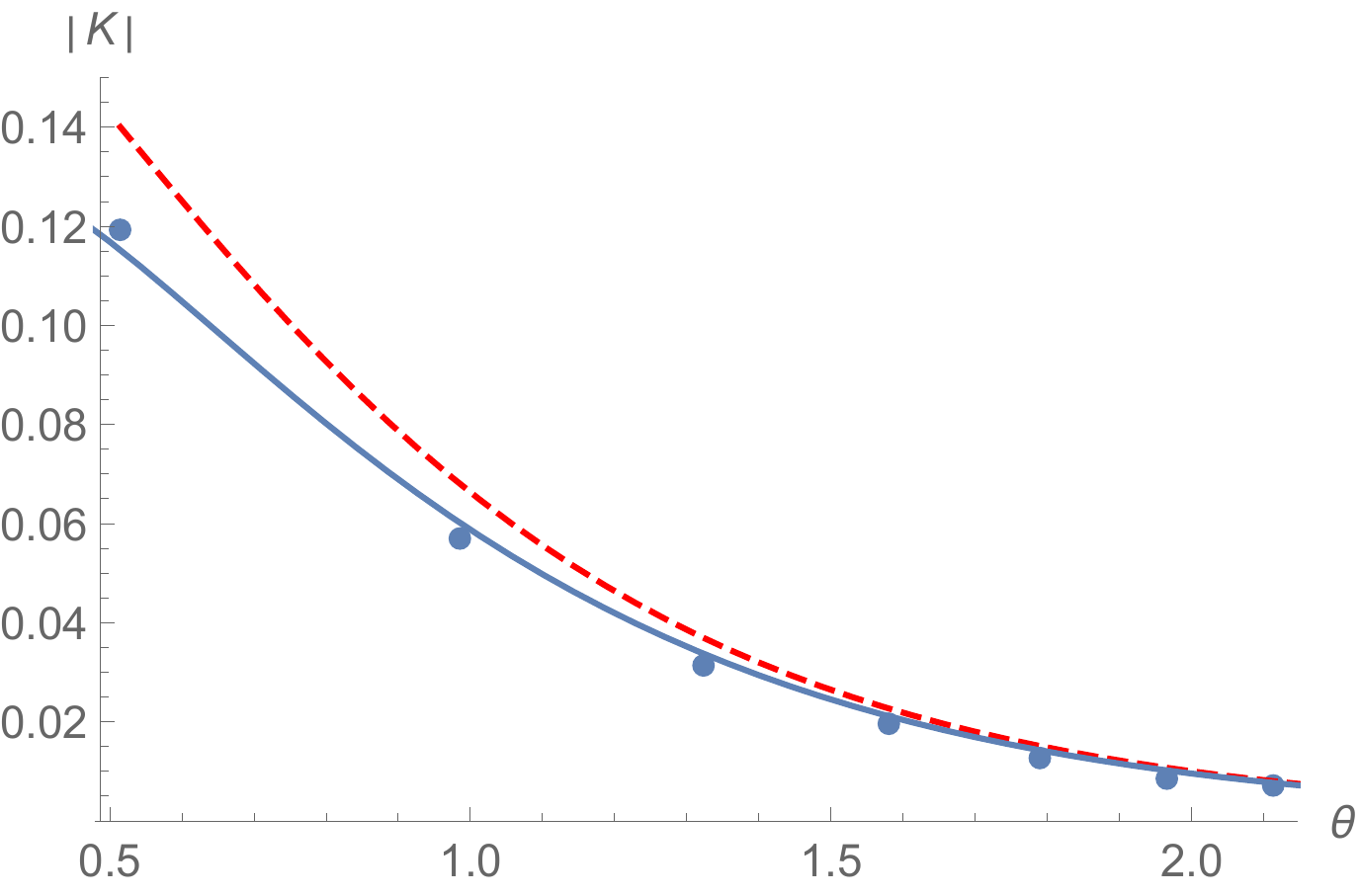}
\par\end{centering}

}\subfloat[$R=2.0$, $M/M_{0}=0.5$, $M_{0}L=50$, $e_{cut}=24$ ]{\begin{centering}
\includegraphics[width=0.45\columnwidth]{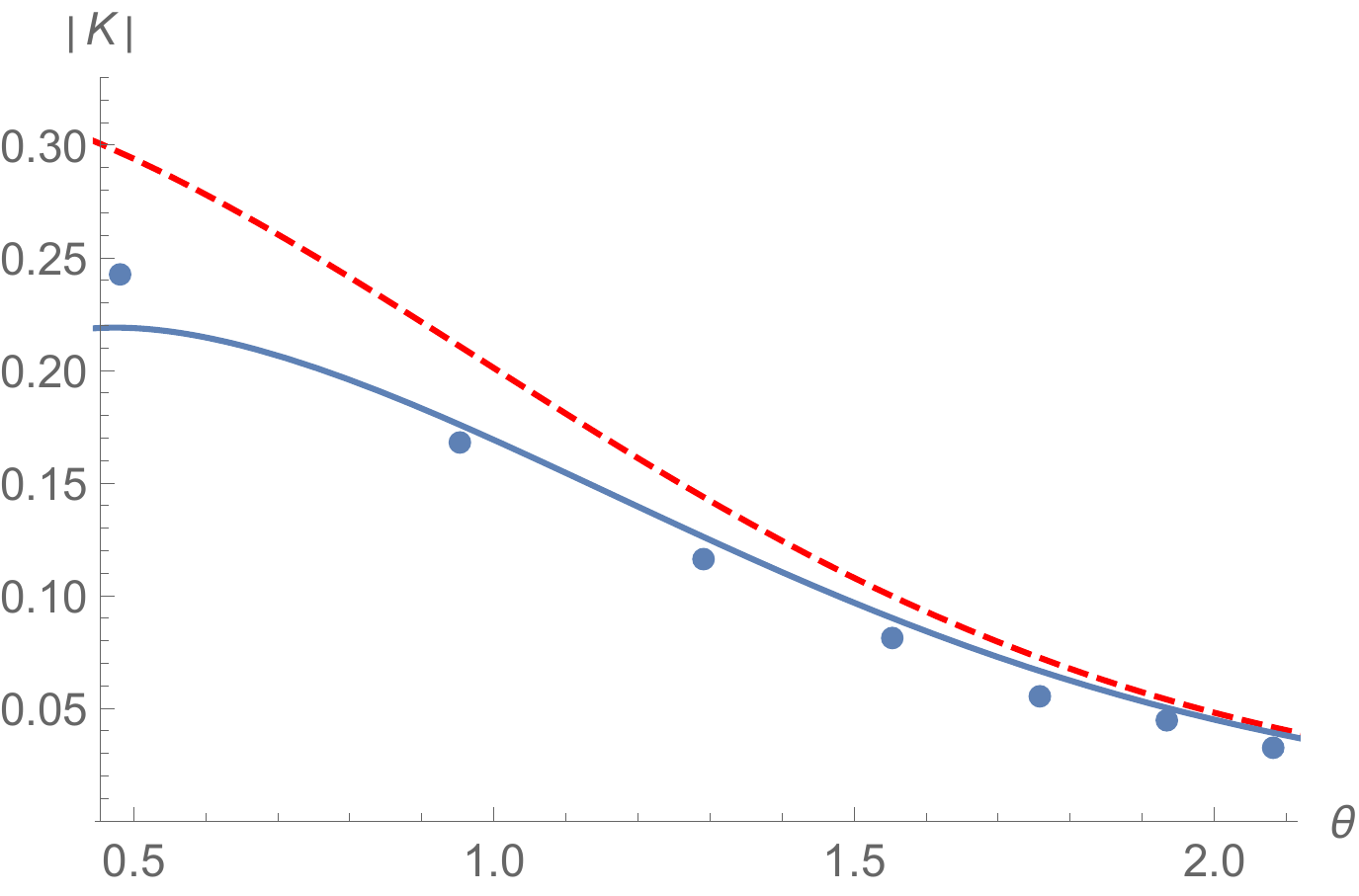}
\par\end{centering}

}

\caption{\label{fig:The-pair-amplitude}The pair amplitude for some mass quenches
in sine-Gordon theory. The sine-Gordon coupling $\beta$ is parametrized
as $\beta=\sqrt{4\pi}/R$. The blue (continuous) curves correspond
to the sine-Gordon Ansatz \eqref{eq:sGAnsatz}, and the red (dashed)
ones to the free theory solutions.}
\end{figure}

Having identified the proper states in the set of numerical eigenstates
the numerical overlaps can be obtained from their scalar product with
the initial state, divided by the vacuum overlap to eliminate the
normalization factor $\mathcal{N}$ in \eqref{eq:ExpInitStateShG}.
As the TCSA matrix elements of the perturbing operator are real numbers,
all the numerically computed eigenvectors are also real, corresponding
to a specific convention for the phases of the post-quench eigenstates.
Therefore the phase of the overlap function $K(\vartheta)$ is absent
from the data, so after normalizing the TCSA overlap values with the
inverse of \eqref{eq:NFunctionsPaired} we compare their modulus to
the value obtained from \eqref{eq:sGAnsatz}. This comparison is shown
in Fig. \ref{fig:The-pair-amplitude} for a few of the quenches we
considered; the conclusion is that it works well except in the low
energy range, and that both the free particle and the Dirichlet parts
of the analytic formula \eqref{eq:sGAnsatz} are important.

Deviations in the low energy range can be attributed to two sources.
First, the initial state is different from the free massive vacuum
for which \eqref{eq:sGAnsatz} (or more precisely, its sinh-Gordon
counterpart \eqref{eq:Ansatz}) was obtained. However, the difference
is the presence of a relevant perturbation in the pre-quench Hamiltonian,
which affects most the low-lying modes due to its relevance. Second,
when modelling the finite size effects in Section \ref{sec:TheoryOfOverlaps}
we used a formalism that neglects exponential corrections in the volume,
which normally affect the lower lying states more. Unfortunately,
it is not easy to separate these effects, and so we cannot say anything
more definite about the low-energy behaviour. However, the analytically
continued solution \eqref{eq:sGAnsatz} definitely provides a good
description of the amplitudes in the mid-to-high energy range.

\subsection{Amplitudes for 4 $B_{1}$ particles and factorization}

{\small{}}
\begin{table}
\begin{centering}
{\small{}}\subfloat[$R=2.3$, $M/M_{0}=0.5$, $M_{0}L=55$, $e_{cut}=24$]{\begin{centering}
{\small{}}%
\begin{tabular}{|c|c|c|c|c|}
\hline 
{\small{}Rapidities $\vartheta_{1}^{*},\vartheta_{2}^{*}$} & {\small{}BY energy} & {\small{}TCSA energy} & {\small{}Normalized overlap } & {\small{}Factorized prediction }\tabularnewline
\hline 
{\small{}\{0.671828, 1.44047\}} & {\small{}1.13089} & {\small{}1.13133} & {\small{}0.0255928} & {\small{}0.0244265}\tabularnewline
\hline 
{\small{}\{0.651971, 1.72849\}} & {\small{}1.34668} & {\small{}1.34742} & {\small{}0.0168602} & {\small{}0.0162507}\tabularnewline
\hline 
{\small{}\{1.19428, 1.70726\}} & {\small{}1.51794} & {\small{}1.51918} & {\small{}0.0108541} & {\small{}0.0108043}\tabularnewline
\hline 
{\small{}\{0.642841, 1.95028\}} & {\small{}1.56712} & {\small{}1.56853} & {\small{}0.0117727} & {\small{}0.0113951}\tabularnewline
\hline 
{\small{}\{0.637471, 2.1315\}} & {\small{}1.73764} & {\small{}1.79245} & {\small{}0.0083998} & {\small{}0.0083603}\tabularnewline
\hline 
\end{tabular}
\par\end{centering}{\small \par}

{\small{}}{\small \par}}
\par\end{centering}{\small \par}

\begin{centering}
{\small{}}\subfloat[$R=2.0$, $M/M_{0}=0.5$, $M_{0}L=50$, $e_{cut}=24$ ]{\begin{centering}
{\small{}}%
\begin{tabular}{|c|c|c|c|c|}
\hline 
{\small{}Rapidities $\vartheta_{1}^{*},\vartheta_{2}^{*}$} & {\small{}BY energy} & {\small{}TCSA energy} & {\small{}Normalized overlap } & {\small{}Factorized prediction }\tabularnewline
\hline 
{\small{}\{0.549607, 1.22608\} } & {\small{}1.33758} & {\small{}1.33916} & {\small{}0.0296001} & {\small{}0.0278547}\tabularnewline
\hline 
{\small{}\{0.524061, 1.51576\} } & {\small{}1.56955} & {\small{}1.57217} & {\small{}0.0139780} & {\small{}0.0195048}\tabularnewline
\hline 
{\small{}\{1.03577, 1.48932\} } & {\small{}1.74274} & {\small{}1.74686} & {\small{}0.0142756} & {\small{}0.0149960}\tabularnewline
\hline 
{\small{}\{0.512645, 1.73741\} } & {\small{}1.80847} & {\small{}1.81308} & {\small{}0.0137404} & {\small{}0.0141252 }\tabularnewline
\hline 
{\small{}\{1.00938, 1.72497\} } & {\small{}1.98019} & {\small{}1.98813} & {\small{}0.0130837} & {\small{}0.0108970}\tabularnewline
\hline 
\end{tabular}
\par\end{centering}{\small \par}

{\small{}}{\small \par}}
\par\end{centering}{\small \par}

\begin{centering}
{\small{}}\subfloat[$R=2.3$, $M/M_{0}=0.75$, $M_{0}L=40$, $e_{cut}=22$]{\begin{centering}
{\small{}}%
\begin{tabular}{|c|c|c|c|c|}
\hline 
{\small{}Rapidities $\vartheta_{1}^{*},\vartheta_{2}^{*}$} & {\small{}BY energy} & {\small{}TCSA energy} & {\small{}Normalized overlap } & {\small{}Factorized prediction }\tabularnewline
\hline 
{\small{}\{0.618879, 1.36023\}} & {\small{}1.60354} & {\small{}1.60489} & {\small{}0.00454695} & {\small{}0.00344512}\tabularnewline
\hline 
{\small{}\{0.599521, 1.64559\} } & {\small{}1.89691} & {\small{}1.89943} & {\small{}0.00292898} & {\small{}0.00219769}\tabularnewline
\hline 
{\small{}\{1.12225, 1.62472\}} & {\small{}2.12314} & {\small{}2.12725} & {\small{}0.00176750} & {\small{}0.00132404}\tabularnewline
\hline 
{\small{}\{0.590723, 1.866\}} & {\small{}2.19785} & {\small{}2.20287} & {\small{}0.00203194} & {\small{}0.00150384}\tabularnewline
\hline 
{\small{}\{1.10091, 1.85633\}} & {\small{}2.42300} & {\small{}2.43139} & {\small{}0.00128401} & {\small{}0.00091002}\tabularnewline
\hline 
\end{tabular}
\par\end{centering}{\small \par}

{\small{}}{\small \par}}
\par\end{centering}{\small \par}

{\small{}\caption{\label{tab:--paired-state}Overlaps for $4$-$B_{1}$ paired states
$|-\vartheta_{1}^{*},\vartheta_{1}^{*},-\vartheta_{2}^{*},\vartheta_{2}^{*}\rangle$.
The sine-Gordon coupling $\beta$ is parametrized as $\beta=\sqrt{4\pi}/R$.
To eliminate differences in phase conventions of energy eigenstates
the modulus of the overlaps is reported.}
}{\small \par}
\end{table}
Once the amplitude $K(\vartheta)$ is pinned down, all higher overlaps
are determined by the exponential form of the state. This entails
the factorisation property which states that states which do not have
an exclusive pair structure in terms of particles have zero overlap,
and for paired states the overlap is just equal to the product of
individual pair state overlaps.

Another prediction from factorization is that the overlaps for paired
$4$-$B_{1}$ states is the product of pair overlaps. This is also
consistent with the TCSA data as shown in Table \ref{tab:--paired-state}.
For the quenches in sub-tables (a) and (b), the overlaps are large
enough so that one can observe a quantitative agreement between the
predictions of \eqref{eq:sGAnsatz} and the TCSA results. For the
example in sub-table (c), the overlap is too small to be measured
and the agreement is only qualitative. 

The question is whether this constitutes a non-trivial test of overlap
factorization? Note that when the quench is small factorization is
expected to be valid to a very good approximation. A small quench
means that the average energy density $\mathcal{E}$ after the quench
satisfies
\[
\mathcal{E}=\frac{1}{L}\left(\langle\Psi(0)|H|\Psi(0)\rangle-\langle0|H|0\rangle\right)\ll m_{1}^{2}
\]
with respect to the mass of the lightest particle $m_{1}$. In such
a case the density of even the lightest pairs is so small that the
average distance $d$ between pairs is much larger than the correlation
length $m_{1}^{-1}$. Since the interactions are suppressed by the
distance as $e^{-m_{1}d}$, the multi-pair amplitudes are expected
to factorize irrespective of integrability when the quench is small. 

We evaluated $\mathcal{E}$ for all the quenches for which we could
produce reliable TCSA data and found that $d$ was at least an order
of magnitude larger than $m_{1}^{-1}$, therefore all observed deviations
from factorization are expected to be TCSA related (either truncation
errors or unmodeled finite size effects). Indeed, when testing the
overlaps for non-paired $4$-$B_{1}$ states, they proved to be an
order of magnitude smaller than the overlaps for paired states, and
were of the same order as the deviations between the prediction \eqref{eq:sGAnsatz}
and the measured two-particle overlap, which is consistent with factorization.

\section{Conclusions \label{sec:Conclusions}}

In this paper we studied mass quenches in the sine-Gordon integrable
quantum field theory in the attractive regime, in particular, we numerically
determined the two-particle overlaps for the $B_{1}$ breathers in
the finite volume theory with a periodic boundary condition. The main
results of our paper is the verification an Ansatz \eqref{eq:Ansatz}
and the exponential form of the initial state \eqref{eq:IntegrableQuench}
proposed in \citep{SotiriadisTakacsMussardo,InitalStateIntEqHierarchcy}
for quenches from the free bosonic theory to the interacting sinh-Gordon
theory. Based on the well-known analytic continuation between the
sine- and sinh-Gordon theories, the numerical overlaps were compared
with the Ansatz. The numerical data points and the theoretical curve
were found to match very well in the middle and high energy range,
with some quantitative deviations in the low energy part which can
be attributed to initial state interactions and finite size effects.
These results confirm the validity of the sinh-Gordon Ansatz, whose
original derivation relied on the assumption that the initial state
contains only multiple particle states composed of pairs with opposite
momenta, which lacks rigorous justification at this moment.

For the numerical determination of the overlaps the truncated conformal
space approach was used. To improve upon the usual renormalization
group treatment of TCSA \citep{TCSARG1,TCSARG2,RGTCSAWatts}, we added
non-local counter terms that dominate the next order corrections in
the inverse energy cut-off. Whereas in general, the construction of
such non-local terms is difficult, in sine-Gordon TCSA it is easy
to implement this correction and the accuracy of the numerical spectrum
can be substantially improved. 

As a closing remark, it has to be mentioned that the quantum sine-Gordon
theory is a very interesting model in its own right, attracting a
lot of attention due to its theoretical tractability and experimental
relevance. Quenches in sine-Gordon theory have recently become realizable
in experimental setups, describing the evolution of the relative phase
of trapped and coupled condensates of cold atoms \citep{Schmiedmayer}.
The knowledge of the pair amplitudes in \eqref{eq:IntegrableQuench}
is crucial for the computation of steady state one- and two-point
functions by currently available techniques \citep{SineGSemiClassical,BertiniSineG},
and the method presented in this paper provides a direct way to the
numerical determination of the pair overlaps. Indeed, in addition
to the $B_{1}$ overlaps presented here, our method can be used to
extract pair amplitudes for higher breathers and soliton-antisoliton
pairs. In this work we refrained from reporting the corresponding
numerical data, since at present we have no theoretical description
for them. The understanding of these overlaps, which is important
for a full description of sine-Gordon quenches, is relegated to future
works.

\subsection*{Acknowledgements}

The authors are grateful to Márton Kormos and Tibor Rakovszky for
their contributions in an early stage of this work and for useful
discussions. This research was supported by the Momentum grant LP2012-50
of the Hungarian Academy of Sciences and by the K2016 grant no. 119204
of the research agency NKFIH.

\end{document}